\documentstyle[a4,12pt]{article}

\newcommand{\ww}{Weizs\"acker-Williams}

\newcommand{\lc}{linear collider}
\newcommand{\sm}{standard model}

\newcommand{\lsp}{lightest supersymmetric particle}
\newcommand{\xs}{cross section}
\newcommand{\br}{branching ratio}

\newcommand{\EW}{electroweak}
\newcommand{\vev}{vacuum expectation value}

\newcommand{\nwa}{narrow width approximation}

\newcommand{\susic}{supersymmetric}
\newcommand{\susy}{supersymmetry}

\newcommand{\sel}{selectron}
\newcommand{\trm}{transverse momentum}

\newcommand{\cm}{center of mass}
\newcommand{\Sel}{\mbox{$\tilde e$}}
\newcommand{\snu}{sneutrino}
\newcommand{\Snu}{\mbox{$\tilde\nu$}}
\newcommand{\co}{chargino}

\newcommand{\no}{neutralino}
\newcommand{\No}{\mbox{$\tilde\chi^0_1$}}
\newcommand{\pT}{\mbox{$p_{\perp}$}}
\newcommand{\mpT}{\mbox{$p_{\perp}\!\!\!\!\!\!/\,\,\,$}}
\newcommand{\ep}{\mbox{$e^+e^-$}}
\newcommand{\ee}{\mbox{$e^-e^-$}}
\newcommand{\eg}{\mbox{$e^-\gamma$}}
\newcommand{\pp}{\mbox{$\gamma\gamma$}}
\newcommand{\GeV}{\mbox{ GeV}}

\newcommand{\hra}{\hspace{1mm}\hookrightarrow}

\def\lr3{$SU(3)_L\otimes SU(3)_R$}

\def\z0{$Z^0$}
\def\Z0{$Z^0$}

\def\ep{$e^+e^-$}

\def\eg{$e^-\gamma$}
\def\cm{centre of mass}

\def\gsim{\buildrel{\lower.7ex\hbox{$>$}}\over{\lower.7ex\hbox{$\sim$}}}
\def\lsim{\buildrel{\lower.7ex\hbox{$<$}}\over{\lower.7ex\hbox{$\sim$}}}

\newenvironment{comment}[1]{}{}

\newcommand{\beq}{\begin{eqnarray}}
\newcommand{\eeq}{\end{eqnarray}}

\begin{document}

\thispagestyle{empty}
\setcounter{page}{0}

\begin{flushright}
MPI-Ph/93-70\\
LMU-93/12\\
September 1993
\end{flushright}
\vspace*{\fill}
\begin{center}
{\Large\bf Discovering Supersymmetry \\
with Electron and Photon Beams\footnote{
Contribution to the Proceedings of the Workshop
on Physics with $e^+e^-$ Linear Colliders,
Annecy, February 15-16, 1993.
}}\\
\vspace{2em}
\large
\begin{tabular}[t]{c}
Frank Cuypers$^{a}$\\
Geert Jan van Oldenborgh$^{b}$\\
Reinhold R\"uckl$^{a,c}$\\
\\
{$^a$ \it Sektion Physik der Universit\"at M\"unchen,
D--80333 M\"unchen, FRG}\\
{$^b$ \it Paul Scherrer Institut, CH-5232 Villigen PSI,
Switzerland}\\
{$^c$ \it Max-Planck-Institut f\"ur Physik,
Werner-Heisenberg-Institut,}\\
{\it D--80805 M\"unchen, FRG}\\
%{$^d$ \it CERN,
%CH-1211 Gen\`eve 23, Switzerland}\\
\end{tabular}
\end{center}
\vspace*{\fill}

\begin{abstract}
\noindent
Linear collider designs are optimized for \ep\ collisions. However,
from the physics point of view, \lc s
may also be advantageously operated in an \ee, \eg\ or \pp\ mode.
These options,
which have not been available up to now,
will provide unique tests of the \sm\ and of physics beyond it.
As an example,
we review the prospects for discovering and studying \susy\
at linear colliders
in \ee, \eg\ and \pp\ collisions.
In particular,
we argue that \ee\ scattering is much better suited
than \ep\ annihilation
for discovering \sel s or \co s.
\end{abstract}
\vspace*{\fill}

\newpage

\section{Introduction}

Linear colliders would be most versatile tools in
experimental high energy physics.
Not only can they provide \ep\ collisions at
high energies and luminosities,
but also very energetic beams of real photons.
One could thus exploit \pp, \eg\ and even \ee\ collisions
for physics studies.
In addition, it appears feasible to produce beams with a
high degree of polarization.
Summing up all combinations of polarizations and beams,
one finds that at least 14 different experiments
can be performed at the same \lc.

These exciting prospects have prompted
a growing number of theoretical studies devoted
to the investigation of the physics potential
of such new accelerator experiments.
Just to cite a few of the possibilities:
\ee\ reactions are ideal for observing lepton number violating processes
\cite{F1,H1};
\eg\ scattering would reveal excited electrons as sharp resonances
\cite{LC29};
\pp\ collisions are powerful probes of the CP quantum numbers
of Higgs bosons \cite{LC27}
and of anomalous gauge couplings \cite{LC28}.

We summarize here the latest results known to us
in the field of \susic\ phenomenology
at \lc s
with polarized and unpolarized
electron and photon colliding beams.
For this
we first outline how high energy photon beams can be obtained
\cite{LC12}.
We then discuss the \susy\ parameters
which are relevant to \sel\ and \co\ production and decay.
In section 4,
we consider the production of \co s and \sel s
in polarized and unpolarized \ee\ collisions \cite{LC11,wo19}.
We emphasize that this channel is almost free from standard model
backgrounds,
in contrast to \ep\ reactions.
Section 5 is
devoted to the study of \eg\ \cite{LC8,LC30,LC19,wo17} collisions, while
section 6 deals with \pp\ \cite{LC25,wo18} collisions.
Finally,
we summarize the main conclusions of this study
and point out questions which require further investigation.

\section{Photon Beams}

High energy photons can be obtained in three ways at linear colliders:
\begin{enumerate}
\item Bremsstrahlung off the initial electrons or positrons.
        The spectrum is given in the \ww\ approximation \cite{WW}
        by the familiar equivalent photon function
        \beq
        P(y) &=& {\alpha\over2\pi} {1+(1-y)^2\over y}
        \ln{s_{ee}\over m_e^2}
        \label{e1}\ ,
        \eeq
        where $\alpha$ is the fine structure constant,
        $y$ is the fraction $E_\gamma/E_e$ of the
        photon and $e^{\pm}$ beam energies,
        $m_e$ is the electron mass, and
        $\sqrt{s_{ee}}=2E_e$ is the nominal collider energy
        in the \ep\ mode.
\item Beamstrahlung
        which results from the interaction of the two beams.
        In general,
        the shape and intensity of the photon distribution
        depend very much on the beam parameters \cite{LC26}.
\item Compton back-scattering of intense laser pulses
        on the bunches of one or both beams.
        The energy distribution
        is given by \cite{LC12}
        \beq
        P(y) &=& {1\over N} \left( 1-y+{1\over1-y}-{4y\over
        x(1-y)}+{4y^2\over x^2(1-y)^2} \right)
        \label{e3}\ ,
        \eeq
        where
        \beq
        0\leq y\leq{x\over x+1}
        &\qquad{\rm and}\qquad&
        x={4E_eE_{\rm laser}\over m_e^2}
        \label{e4}\ .
        \eeq
        The factor
        \beq
        N &=& {1\over2} + {8\over x} - {1\over2(1+x)^2} +
        \left( 1-{4\over x}-{8\over x^2} \right) \ln(1+x)
        \label{e5}
        \eeq
	normalizes $\int dy \,P(y)$ to 1.
\end{enumerate}
In contrast to the first two mechanisms
which take place more or less naturally,
%(beamstrahlung, though, can be strongly increased
%for some beam shapes),
it is not a trivial task to implement laser Compton back-scattering.
On the other hand,
this option yields a much harder photon spectrum
and is thus much better suited for the photoproduction
of heavy particles.
For this reason
we concentrate here on this third option
where the photons obey the distribution (\ref{e3}).

\begin{figure}[hbt]
\begin{center}
\begin{picture}(626,400)(0,0)
{\huge\tt
	The figures can be obtained via anonymous ftp from 129.187.198.1
	in preprints/lc93/susy.ps .
}
\begin{comment}{
\put(0,0){\strut\bBoxedEPSF{spectrum.ps}}
\put(100.8,388.1){\makebox(0,0)[br]{$P(y)$}}
\put(552.9, 19.1){\makebox(0,0)[tr]{$y=E_\gamma/E_{e}$}}
\put(504.6,347.8){\makebox(0,0)[r]{\small back-scattering}}
\put(504.6,327.8){\makebox(0,0)[r]{\small Bremsstrahlung}}
}\end{comment}\end{picture}
\\
\end{center}
\caption[f1]{\small Energy spectra of
photons from laser Compton
back-scattering (\ref{e3}) and Bremsstrahlung (\ref{e1})
for a 250 GeV electron beam.}
\label{f1}
\end{figure}

In what follows,
we assume a 100\%\ conversion efficiency
and neglect the angular dispersion of the back-scattered photons.
When $x$ reaches the value $2(\sqrt{2}+1)\approx4.83$,
the back-scattered and laser photons have enough relative energy
to produce $e^+e^-$ pairs.
As a consequence, the conversion efficiency drops considerably
for larger values of $x$.
We therefore assume the laser energy to be tuned
in such a way as to obtain
$x=2(\sqrt{2}+1)$.
The corresponding energy spectrum (\ref{e3})
is displayed in Fig.~\ref{f1}.
It sharply contrasts
with the energy spectrum of Bremsstrahlung photons,
which is much softer and, hence,
less useful for heavy particle searches.

If the electron and/or laser beams are polarized
the spectrum depicted in Fig.~\ref{f1}
can be substantially modified
and become either softer or harder.
Nevertheless,
the hardest photons obtained in this way
will never exceed an energy of
${x\over x+1} E_e = 2(\sqrt{2}-1) E_e \approx .83 E_e$.
This means that
in \eg\ and \pp\ collisions
the effective \cm\ energy is reduced by about 10 \%\ and 20 \%,
respectively,
in comparison to the \ep\ and \ee\ collision energies.

Moreover, the \xs s of \eg\ and \pp\ reactions
involving such a back-scattered photon beam
have to be folded with the energy distribution (\ref{e3}).
The laboratory frame is thus not the \cm\ frame
and the \eg\ and $\gamma\gamma$ \cm\ energies
$\sqrt{s_{e\gamma}}$ and $\sqrt{s_{\gamma\gamma}}$, respectively,
are given by
\beq
s_{e\gamma}=ys_{ee}
&\qquad{\rm and}\qquad&
s_{\gamma\gamma}=y_1y_2s_{ee}\ ,
\eeq
where the different $y$'s denote the photon energy fractions.
Finally, the convoluted \xs s are obtained from
\beq
\sigma(s_{ee})=\int \!\!dy\ P(y)\sigma(s_{e\gamma})
&\quad{\rm and}\quad&
\sigma(s_{ee})=\int \!\!dy_1\int
        \!\!dy_2\ P(y_1)P(y_2)\sigma(s_{\gamma\gamma})\ .
\eeq

\section{Supersymmetry Parameters}

In spite of its economy of principles,
a \susic\ extension of the standard model \cite{HKN}\
involves an opulent number of additional free parameters.
Although the strengths of the interactions
must be precisely the same as those of the \sm,
at this stage
the masses and eventual mixings of the \susic\ partners
of the conventional particles
cannot be predicted from first principles.
The most relevant parameters here
are the mass parameters $M_1, M_2, \mu$ associated with
the $U(1)$ and $SU(2)_L$ gauginos and the higgsinos respectively,
the ratio $\tan\beta=v_2/v_1$ of the Higgs \vev s
and the slepton masses.
We work in the context of the
minimal \susic\ standard model
and make the following assumptions:
\begin{enumerate}
\item   R-parity is a conserved quantum number.
\item   The lightest neutralino $\tilde\chi^0_1$ is the \lsp.
\item   All sleptons have the same mass
        and are much lighter than the strongly interacting
        squarks and gluinos:
        $m_{\tilde\ell_L}=m_{\tilde\ell_R}=m_{\Snu_{\ell}}
        \ll m_{\tilde q},m_{\tilde g}$.
\item   The mass parameters $M_1, M_2, \mu$ are real.
\item   At the GUT scale $M_1=M_2$,
        so that after renormalization to accelerator energies
        $M_1=5/3\,M_2\tan^2\theta_w$,
        where $\theta_w$ is the weak mixing angle.
\end{enumerate}
The first two assumptions are essential,
because they dictate the whole \susic\ phenomenology:
all sparticles decay directly or via a cascade
into the \lsp\
which is stable and escapes detection.
The last three assumptions are made for simplicity
and can be relaxed without modifying qualitatively the conclusions.
As a consequence of these assumptions
there are thus only four parameters left

\medskip
\centerline{\fbox{$\tan\beta \quad M_2 \quad \mu \quad m_{\tilde\ell}$}}

which are relevant for the present studies.
Of these four parameters,
$\tan\beta$ is the least influential,
at least when it is larger than 2.
In contrast,
the results can be very sensitive
to variations of the other three parameters.

In the following
we only consider the production of the lightest \co\ and sleptons
(typically the \sel),
since according to our third assumption
strongly interacting sparticles are too heavy to be produced competitively.
Since the \co\ and the \sel\
decay only through \EW\ interactions,
their lifetimes are typically long
in comparison to their mass scale.
It is therefore
safe to use the \nwa.

The simplest decay mode of the \sel\
is into an electron and the lightest \no:
\begin{equation}
        \tilde e^-\to e^-\No~.
\label{seldec}
\end{equation}
Since we assume the \no\ to be the \lsp,
only the electron is visible.
If kinematically allowed,
other decays can take place like
\begin{eqnarray}
\tilde e^-
&\to&
e^-\tilde\chi^0_2~,
\label{seldec1}\\
%\tilde e^-
&\to&
\nu_e\tilde\chi^-_1~,
\label{seldec2}
\end{eqnarray}
and similar decays into the heavier neutralino and chargino states.
The \susic\ particles produced in this way
will decay into lighter (s)particles,
which themselves might undergo further decays
until only conventional particles and a number of
the lightest \no\ remain.
The end-product of such cascade decays
can sometimes again be an electron
accompanied by invisible particles only.
Whenever such cascade decays are important
({\em i.e.} when no high \trm\ cuts
need to be imposed on the emerging electrons)
we compute the \br\ for the decay
$\tilde e^-\to e^-+{\rm invisible}$,
with the two-body decay algorithm
described in Ref.~\cite{wo22}.
Typically,
the left-\sel\
({\em i.e.} the partner of the left-handed electron)
has a lower \br\ for the decay (\ref{seldec})
than the right-\sel,
because the latter cannot decay into \co s.

The decays of \co s are more complicated.
Concentrating on the lightest \co,
if kinematically allowed to do so,
it will decay into leptons and sleptons or charged bosons and \no s:
\begin{eqnarray}
\tilde\chi^-_1
&\to&
\ell^-\Snu_\ell~,
\label{codec1}
\\
&\to&
\tilde\ell^-\nu_\ell~,
\label{codec2}
\\
&\to&
W^-\tilde\chi_i^0~,
\label{codec3}
\\
&\to&
H^-\tilde\chi_i^0
\label{codec4}\ .
\end{eqnarray}
For simplicity,
we discard here the last possibility (\ref{codec4})
by assuming the charged Higgs boson to be heavy.
If the \co\ is heavier than the sleptons,
it will preferentially decay
with approximately a 50\%\ \br\ in each of the channels
(\ref{codec1}) and (\ref{codec2})
and with democratic
probabilities for the different flavours \cite{wo18}.
In this case,
the sleptons can only decay further
into leptons and \no s.
If the \co\ is lighter than the sleptons
but can still decay according to the reaction (\ref{codec3}),
one has to deal with a $W^-$ signal.
It can also happen that none of the two-body decays
(\ref{codec1}-\ref{codec4})
is kinematically allowed.
If this is the case
and if the mass of the sleptons is much larger than the mass of the $W$,
the decay through a virtual $W$ dominates.
The \br\ of the leptonic decay
$\tilde\chi^-_1\to\ell^-\bar\nu_\ell\No$
is then approximately 42\%\
\cite{LC14}.

In the next sections,
where we consider \sel\ and \co\ production,
we focus on the decay signatures:

\centerline{\fbox{$
        \begin{array}{ccc}
        \tilde e^- &\to& e^- + \mpT
        \\
        \tilde \chi^-_1 &\to& \mu^- + \mpT
        \end{array}
$}}

\section{$e^-e^-$ Collisions}

In $e^-e^-$ collisions
\co\ and \sel\ production are both accessible.
The most striking signatures are, respectively,
$e^-e^-\to\mu^-\mu^-+\mpT$ and $e^-e^-\to e^-e^-+\mpT$.
We summarize here the results \cite{wo19}
for each reaction.

\begin{figure}[htb]
\begin{center}
\begin{picture}(250,500)(0,-200)
{\tt
	The figures can be obtained via anonymous ftp from 129.187.198.1
	in preprints/lc93/susy.ps .
}
\begin{comment}{
\ArrowLine(00,00)(37.5,5)
\ArrowLine(0,50)(37.5,45)
\Vertex(37.5,45){.5}
\DashLine(37.5,45)(37.5,5){5}
\Vertex(37.5,5){.5}
\ArrowLine(37.5,5)(75,0)
\ArrowLine(37.5,45)(75,50)
\Text( -5,0)[r]{$e^-$}
\Text( -5,100)[r]{$e^-$}
\Text( 80,50)[l]{$\tilde\nu$}
\Text(155,100)[l]{$\tilde{\chi}_1^-$}
\Text(155,0)[l]{$\tilde{\chi}_1^-$}
\Text(30,-50)[l]{+ crossed}
}\end{comment}\end{picture}
%\begin{picture}(554,504)(0,0)
\begin{picture}(554,300)(0,0)
\begin{comment}{
\put(0,0){\strut\bBoxedEPSF{/user/frank/Papeles/LC93/co.ps}}
\Text(65,460)[tr]{\small$\sigma$[pb]}
\Text(520,19)[tr]{\small$\sqrt{s_{ee}}$[GeV]}
\Text(300,450)[tl]
{\fbox{\small $e^-e^- \to \tilde\chi_1^-\tilde\chi_1^-$}}
\Text(300,375)[tl]{\small $\tan\beta = 10$}
\Text(300,350)[tl]{\small $\mu = -300$ GeV}
\Text(300,325)[tl]{\small $M_2 = 200$ GeV}
\Text(300,300)[tl]{\small $m_{\tilde\ell} = 150$ GeV}
}\end{comment}\end{picture}
\end{center}
\caption[f2]{\small Lowest order Feynman diagram
describing \co\ production in $e^-e^-$ collisions
and typical energy dependence of the corresponding \xs.
The values chosen for the \susic\ parameters imply
$m_{\tilde\chi_1^-}= 186$ GeV.}
\label{f2}
\end{figure}

\subsection{Chargino Production}

Chargino production takes place in $e^-e^-$ collisions
via the exchange of a sneutrino,
as depicted in Fig.~\ref{f2}.
The energy dependence of the \xs\
is also shown in Fig.~\ref{f2}
for a particular choice of \susy\ parameters.
The yield is sharply peaked just above threshold
so that an appropriate
energy scan can provide information about the mass of the \co.
We focus here on the following observable signal
originating from the chargino
decays (\ref{codec1}--\ref{codec3})

\centerline{\fbox{$\ee \to \tilde\chi_1^-\tilde\chi_1^- \to \mu^-\mu^-+\mpT$}}

The \xs\ for obtaining this signal
can be much lower than the total production \xs,
because the \br\ of the decay of a \co\ into a muon and
invisible particles is always lower than 1/3.
Still,
the dominant \sm\ background due to the process
{\arraycolsep0cm
\renewcommand{\arraystretch}{0}
\begin{equation}
        e^-e^- \to
        \begin{array}[t]{ll}
                W^-\nu_e&W^-\nu_e \qquad ,\\
                        &\hra\mu^-\bar\nu_\mu\\
                \multicolumn{2}{l}{\hra\mu^-\bar\nu_\mu}
        \end{array}
\label{ww}
\end{equation}
}
should not exceed 1 fb
for $\sqrt{s_{ee}}=500$ GeV.
This is very different in \ep\ reactions,
where it is difficult to disentangle
a \susic\ $\mu^+\mu^-+\mpT$ signal
from the large \sm\ background,
mainly due to $W$ pair production.

In Fig.~\ref{f4} we have plotted
the contours in the $(\mu,M_2)$ half-plane
along which the observable \xs\ for the $\mu^-\mu^-+\mpT$ signal
from the decay modes (\ref{codec1}-\ref{codec3})
is 1, 10 and 100 fb.
For this we assumed a mass of 300 GeV for the exchanged \snu.
As long as $M_2\lsim300$ GeV,
the signal to background ratio should comfortably exceed one.
To compute these \xs s
we used a theorists' detector
by imposing the following rapidity, energy and acoplanarity cuts
on the observed leptons:
\begin{equation}
|\eta_e| < 3
\quad , \quad
E_e > 5 \GeV
\quad , \quad
||\phi(\ell^-_1)-\phi(\ell^-_2)|-180^\circ| > 2^\circ
\label{cut}\ ,
\end{equation}
where $\phi$ is the azimuthal angle of the decay leptons with
respect to the beam axis.

\begin{figure}[htb]
\centerline{
\begin{picture}(250,500)(0,-350)
{\tt
	The figures can be obtained via anonymous ftp from 129.187.198.1
	in preprints/lc93/susy.ps .
}
\begin{comment}{
\Text(0,0)[tl]{\small \fbox{$e^-e^- \to \mu^-\mu^-+\mpT$}}
\Text(0,-75)[tl]{\small $\sqrt{s_{ee}} = 500$ GeV}
\Text(0,-100)[tl]{\small $\tan\beta = 10$}
\Text(0,-125)[tl]{\small $m_{\tilde\ell} = 300$ GeV}
}\end{comment}\end{picture}
\begin{picture}(554,504)(0,0)
\begin{comment}{
\put(0,0){\strut\bBoxedEPSF{cc.hacked.ps}}
\Text(70,460)[tr]{\small$M_2$[GeV]}
\Text(481,19)[tr]{\small$\mu$[GeV]}
\Text(291,133)[b]{\small LEP2}
\Text(177,217)[l]{\small 100 fb}
\Text(196,309)[l]{\small 10 fb}
\Text(215,406)[l]{\small 1 fb}
\Text(148,406)[b]{\scriptsize no pair-}
\Text(148,377)[b]{\scriptsize production}
\Text(433,406)[b]{\scriptsize no pair-}
\Text(433,377)[b]{\scriptsize production}
}\end{comment}\end{picture}
}
\caption[f4]{\small Contours in the \susy\ parameter space
of constant \xs s for the chargino signal.
The cuts (\ref{cut}) are included.
The chargino mass varies with $M_2$ and $\mu$.}
\label{f4}
\end{figure}

\subsection{Selectron Production}

Selectron production takes place in \ee\ collisions
via the exchange of \no s,
as depicted in Fig.~\ref{f5}.
Note that all four \no s play an important role in this reaction
\cite{wo22}.
The dependence on the \susy\ parameters is thus rather complex
because it enters at three different levels:
(i) through the masses of the four different \no s;
(ii) through their mixings among each other which affects
their couplings to electrons and \sel s;
(iii) through the mass and \br s of the \sel.
%A single \ee\ experiment can thus not provide a determination
%of the \susy\ parameters.

\begin{figure}[htb]
\begin{center}
\begin{picture}(250,500)(0,-200)
{\tt
	The figures can be obtained via anonymous ftp from 129.187.198.1
	in preprints/lc93/susy.ps .
}
\begin{comment}{
\ArrowLine(00,00)(37.5,5)
\ArrowLine(0,50)(37.5,45)
\Vertex(37.5,45){.5}
\Line(37.5,45)(37.5,5)
\Vertex(37.5,5){.5}
\DashLine(37.5,5)(75,0){5}
\DashLine(37.5,45)(75,50){5}
\Text( -5,0)[r]{$e^-$}
\Text( -5,100)[r]{$e^-$}
\Text( 80,50)[l]{$\tilde\chi^0_i$}
\Text(155,100)[l]{$\tilde{e}^-$}
\Text(155,0)[l]{$\tilde{e}^-$}
\Text(30,-50)[l]{+ crossed}
}\end{comment}\end{picture}
\begin{picture}(554,504)(0,0)
\begin{comment}{
\put(0,0){\strut\bBoxedEPSF{/user/frank/Papeles/LC93/ss.ps}}
\Text(70,460)[tr]{\small$\sigma$[pb]}
\Text(520,19)[tr]{\small$\sqrt{s_{ee}}$[GeV]}
\Text(300,450)[tl]{\fbox{\small $e^-e^- \to \tilde e^-\tilde e^-$}}
\Text(300,375)[tl]{\small $\tan\beta = 10$}
\Text(300,350)[tl]{\small $\mu = -300$ GeV}
\Text(300,325)[tl]{\small $M_2 = 200$ GeV}
\Text(300,300)[tl]{\small $m_{\tilde\ell} = 150$ GeV}
}\end{comment}\end{picture}
\end{center}
\caption[f5]{\small Lowest order Feynman diagram
describing \sel\ production in $e^-e^-$ collisions
and typical energy dependence of the corresponding \xs.}
\label{f5}
\end{figure}

The energy dependence of the \xs\
is shown in Fig.~\ref{f5}
for the same choice of \susy\ parameters as previously.
Also here, the yield is sharply peaked just above threshold
so that an energy scan can provide precise information
about the mass of the \sel.
The prominent decay mode (\ref{seldec})
leads to the following observable signal:

\medskip
\centerline{\fbox{$\ee \to \tilde e^-\tilde e^- \to \ee+\mpT$}}

The \xs\ for obtaining this signal
is comparable with the total production \xs\ in a large part of the
parameter space considered.
Indeed,
if the \sel\ is light
it can only decay as in Eq.~\ref{seldec}
so that the \br\ of the decay of a \sel\ into an electron and invisible
particles is one.
In addition,
for most choices of the relevant \susy\ parameters
the right-\sel\ will exclusively decay according to Eq.~\ref{seldec}.
The most important background from the \sm\ processes
{\arraycolsep0cm
\renewcommand{\arraystretch}{0}
\beq
        e^-e^- \to
        \begin{array}[t]{ll}
                e^-\nu_e&W^- \qquad \\
                      &\hra e^-\bar\nu_e
        \end{array}
&\qquad{\rm and}\qquad&
        e^-e^- \to
        \begin{array}[t]{ll}
                e^-e^-&Z^0 \qquad \\
                      &\hra\nu\bar\nu
        \end{array}
\label{wz}
\eeq
}
is not particularly high.
After imposing the acceptance cuts (\ref{cut})
and including the relevant branching ratios,
the \xs s are 150 fb for $W^-$
and 40 fb for $Z^0$ Bremsstrahlung,
at $\sqrt{s_{ee}}=500$ GeV.
The potential background from M\o ller scattering
is eliminated by the acoplanarity cut.
The \susic\ signal,
on the other hand,
is not significantly reduced by these mild cuts
(which basically simulate a typical detector acceptance).

Note that softer electrons emerging at the end
of a longer cascade such as the ones initiated by the
decays (\ref{seldec1},\ref{seldec2})
will not be very much affected by the cuts (\ref{cut}) either.
In contrast to the situation in
\ep, \eg\ and \pp\ collisions,
where high \trm\ cuts are necessary in order
to enhance the signal to background ratio,
\ee\ scattering is thus ideal for
observing and studying \susic\ cascades.

In Fig.~\ref{f7} we have plotted
the contours in the $(\mu,M_2)$ half-plane
along which the observable \xs\ for the $e^-e^-+\mpT$ signal
from the production of 200 GeV \sel s
is 1 and 0.1 pb.
The dotted lines
show what is obtained if one ignores
those cascade decays
({\em e.g.} (\ref{seldec1},\ref{seldec2}))
of the \sel\
which yield the same $e^-e^-+\mpT$ final states as the direct
decay (\ref{seldec}).

\begin{figure}[htb]
\centerline{
\begin{picture}(250,500)(0,-350)
{\tt
	The figures can be obtained via anonymous ftp from 129.187.198.1
	in preprints/lc93/susy.ps .
}
\begin{comment}{
\Text(0,0)[tl]{\small \fbox{$e^-e^- \to e^-e^-+\mpT$}}
\Text(0,-75)[tl]{\small $\sqrt{s_{ee}} = 500$ GeV}
\Text(0,-100)[tl]{\small $\tan\beta = 10$}
\Text(0,-125)[tl]{\small $m_{\tilde\ell} = 200$ GeV}
}\end{comment}\end{picture}
\begin{picture}(554,504)(0,0)
\begin{comment}{
\put(0,0){\strut\bBoxedEPSF{np.hacked.ps}}
\Text(70,460)[tr]{\small$M_2$[GeV]}
\Text(481,19)[tr]{\small$\mu$[GeV]}
\Text(291,133)[b]{\small LEP2}
\Text(112.2,426.9)[lb]{\scriptsize unphysical}
\Text(389.6,426.9)[lb]{\scriptsize unphysical}
\Text(116,293)[lb]{\small 1 pb}
\Text(475,293)[rb]{\small 1 pb}
\Text(363,217)[lb]{\small 0.1 pb}
}\end{comment}\end{picture}
}
\caption[f7]{\small Contours in the \susy\ parameter space
of constant \xs s for the selectron signal.
The cuts (\ref{cut}) are included.
The regions labeled ``unphysical'' are excluded since there
$m_{\tilde e} < m_{\tilde\chi^0_1}$.
The contours which would be obtained if cascade decays are ignored
are also shown with dotted lines.
}
\label{f7}
\end{figure}

\subsection{Polarized Beams}

Polarization can enhance the \co\ signal
together with the background
by up to a factor four.
This is a welcome but not dramatic effect.

For \sel s,
on the other hand,
the signal can be strongly enhanced or suppressed with polarized beams,
depending on the values taken by the \susy\ parameters.
The main virtue of polarization,
however,
is that with right-handed electron beams
the $W^-$ Bremsstrahlung background (\ref{wz}) disappears.
It is then worthwhile to also eliminate the background
from $Z^0$ Bremsstrahlung,
in order to select a really clean sample of \susic\ events
with no,
or negligibly little,
background from \sm\ processes.
This can be done by rejecting all \ee\ events
with a total deposited energy exceeding about half the \cm\ energy
\begin{equation}
E_{e_1}+E_{e_2} < {s-m_Z^2\over2\sqrt{s}} \approx 242 \GeV
\label{e1e2cut}\ .
\end{equation}
If this cut is imposed,
none of the $Z^0$ contributes
and at worst 55\%\ of the electron pairs
which originate from \sel\ production are lost.
The dominant irreducible background then
originates from double $W^-$ Bremsstrahlung (\ref{ww}).
It should not amount to more than 1 fb.
Note that with \ep\ beams,
the \sm\ backgrounds cannot be eliminated in such a simple way.
We have plotted in Fig.~\ref{f8}
the same contours as in Fig.~\ref{f7}
for right-polarized beams.

\begin{figure}[htb]
\centerline{
\begin{picture}(250,500)(0,-350)
{\tt
	The figures can be obtained via anonymous ftp from 129.187.198.1
	in preprints/lc93/susy.ps .
}
\begin{comment}{
\Text(0,0)[tl]{\small \fbox{$e_R^-e_R^- \to e^-e^-+\mpT$}}
\Text(0,-75)[tl]{\small $\sqrt{s_{ee}} = 500$ GeV}
\Text(0,-100)[tl]{\small $\tan\beta = 10$}
\Text(0,-125)[tl]{\small $m_{\tilde\ell} = 200$ GeV}
}\end{comment}\end{picture}
\begin{picture}(554,504)(0,0)
\begin{comment}{
\put(0,0){\strut\bBoxedEPSF{rr.hacked.ps}}
\Text(65,460)[tr]{\small$M_2$[GeV]}
\Text(481,19)[tr]{\small$\mu$[GeV]}
\Text(291,133)[b]{\small LEP2}
\Text(112.2,426.9)[lb]{\scriptsize unphysical}
\Text(389.6,426.9)[lb]{\scriptsize unphysical}
\Text(207,259)[r]{\small 1 pb}
\Text(386,259)[l]{\small 1 pb}
\Text(222,452)[l]{\small 1 pb}
\Text(359,452)[r]{\small 1 pb}
\Text(239,172)[r]{\small 0.1 pb}
\Text(352,182)[l]{\small 0.1 pb}
\Text(291,325)[b]{\small 0.1 pb}
}\end{comment}\end{picture}
}
\caption[f8]{\small Same as Fig.~\ref{f7},
for right-handed electron beams
and the additional energy cut (\ref{e1e2cut}).
}
\label{f8}
\end{figure}

Note that with such a polarization experiment,
not only are the allowed values of the \susy\ parameters further restricted,
but also the mass of the \no\ can be kinematically determined
from the endpoints $E_{\rm min,max}$
of the electron energy distribution:
\beq
m_{\tilde\chi^0_1}^2
&=&
\sqrt{s_{ee}}
{E_{\rm max}E_{\rm min} \over E_{\rm max}+E_{\rm min}}
\left( {\sqrt{s_{ee}} \over E_{\rm max}+E_{\rm min}} -2 \right)
\label{e100}\ .
\eeq
Of course,
there will always be some smearing due to initial state Bremsstrahlung
and beamstrahlung.
The incidence of these effects should be further investigated.
We emphasize
%that even if positron beams could be
%polarized by the time the \lc\ is built,
that this interesting possibility does not exist
in \ep\ collisions.

\section{$e^-\gamma$ Collisions}

We have seen in the previous section that
if \sel s or \co s are
light enough to be pair-produced in \ee\ collisions
this will yield an unmistakable signal.
However,
if the collider energy is below the pair-production threshold
it is the \eg\ operating mode which may save the day \cite{wo17}.
As depicted in Fig.~\ref{f9}, \eg\ collisions can produce
selectrons in association with the lightest neutralino and
thus probe higher selectron masses than \ee\ collisions.

\begin{figure}[htb]
\begin{center}
\begin{picture}(250,500)(0,-100)
{\tt
	The figures can be obtained via anonymous ftp from 129.187.198.1
	in preprints/lc93/susy.ps .
}
\begin{comment}{
\ArrowLine(00,100)(25,125)
\Photon(0,150)(25,125){2}{5}
\Vertex(25,125){.5}
\ArrowLine(25,125)(50,125)
\Vertex(50,125){.5}
\Line(50,125)(75,100)
\DashLine(50,125)(75,150){5}
\Text( -5,200)[r]{$e^-$}
\Text( -5,300)[r]{$\gamma$}
\Text( 80,270)[c]{$e^-$}
\Text(155,200)[l]{$\tilde{\chi}^0_1$}
\Text(155,300)[l]{$\tilde{e}^-$}
\Text(75,150)[c]{+}
\ArrowLine(00,00)(37.5,5)
\Photon(00,50)(37.5,45){2}{5}
\Vertex(37.5,45){.5}
\DashLine(37.5,45)(37.5,5){5}
\Vertex(37.5,5){.5}
\Line(37.5,5)(75,0)
\DashLine(37.5,45)(75,50){5}
\Text(-5,0)[r]{$e^-$}
\Text(-5,100)[r]{$\gamma$}
\Text(80,50)[l]{$\tilde e^-$}
\Text(155,100)[l]{$\tilde e^-$}
\Text(155,0)[l]{$\tilde{\chi}_1^0$}
}\end{comment}\end{picture}
\begin{picture}(554,504)(0,0)
\begin{comment}{
\put(0,0){\strut\bBoxedEPSF{/user/frank/Papeles/LC93/eg.ps}}
\Text(70,450)[tr]{\small$\sigma$[pb]}
\Text(520,19)[tr]{\small$\sqrt{s_{ee}}$[GeV]}
\Text(350,450)[tl]{\fbox{\small $e^-\gamma \to \tilde e^-\tilde\chi_1^0$}}
\Text(350,375)[tl]{\small $\tan\beta = 10$}
\Text(350,350)[tl]{\small $\mu = -300$ GeV}
\Text(350,325)[tl]{\small $M_2 = 200$ GeV}
\Text(350,300)[tl]{\small $m_{\tilde\ell} = 150$ GeV}
}\end{comment}\end{picture}
\end{center}
\caption[f9]{\small Lowest order Feynman diagrams
describing \sel\ and \no\ production in $e^-\gamma$ collisions
and typical energy dependence of the corresponding \xs.
For the photon the energy spectrum (\ref{e3}) is assumed.}
\label{f9}
\end{figure}

The energy dependence of the \xs\
is also shown in Fig.~\ref{f9}
for the same choice of \susy\ parameters as previously.
If the \sel\ decays in the channel (\ref{seldec}),
the observed signal is quite striking:

\medskip
\centerline{\fbox{$e^-\gamma\to \tilde e^-\tilde\chi_1^0 \to e^-+\mpT$}}

Unfortunately,
the \sm\ backgrounds from the reactions
{\arraycolsep0cm
\renewcommand{\arraystretch}{0}
\beq
        e^-\gamma \to
        \begin{array}[t]{ll}
                 \nu_e&W^- \qquad \\
                      &\hra e^-\bar\nu_e
        \end{array}
&\qquad{\rm and}\qquad&
        e^-\gamma \to
        \begin{array}[t]{ll}
                   e^-&Z^0 \qquad ,\\
                      &\hra\nu\bar\nu
        \end{array}
\label{ez}
\eeq
}
are substantial.
At $\sqrt{s_{ee}}=500$ GeV
the corresponding \xs s
amount to about 3 pb each.
However,
by imposing the transverse momentum and rapidity cuts
\beq
p_\perp(e^-)>50\ \GeV\
&\qquad{\rm and}\qquad&
0<\eta(e^-)<2
\label{egcuts}\ ,
\eeq
the combined background can be reduced to less than 0.3 pb.
These cuts enhance the signal to background ratio
by more than an order of magnitude.

For the sake of illustration,
we now consider a \sel\ with $m_{\tilde e} = 250$ GeV,
which cannot be pair-produced in 500 GeV $e^\pm e^-$ collisions.
In Fig.~\ref{f11} we show
the contours in the $(\mu,M_2)$ half-plane
along which the signal
exceeds the backgrounds statistical fluctuations
by more than 3 standard deviations, i.e.
$n_{SUSY}>3\sqrt{n_{SM}}$.
The shaded area
shows the part of parameter space
which will already have been investigated
by a \co\ search in $e^\pm e^-$ collisions.
Clearly,
a deeper exploration of the parameter space requires
very high luminosities.

\begin{figure}[htb]
\centerline{
\begin{picture}(250,500)(0,-350)
{\tt
	The figures can be obtained via anonymous ftp from 129.187.198.1
	in preprints/lc93/susy.ps .
}
\begin{comment}{
\Text(0,0)[tl]{\small \fbox{$e^-\gamma \to e^-+\mpT$}}
\Text(0,-75)[tl]{\small $\sqrt{s_{ee}} = 500$ GeV}
\Text(0,-100)[tl]{\small $\tan\beta = 10$}
\Text(0,-125)[tl]{\small $m_{\tilde\ell} = 250$ GeV}
}\end{comment}\end{picture}
\begin{picture}(554,504)(0,0)
\begin{comment}{
\put(0,0){\strut\bBoxedEPSF{/user/frank/Papeles/LC93/eg.lum.ps}}
\Text(65,460)[tr]{\small$M_2$[GeV]}
\Text(520,19)[tr]{\small$\mu$[GeV]}
\Text(310,250)[bc]{\small 100 fb$^{-1}$}
\Text(220,200)[tr]{\small 10 fb$^{-1}$}
\Text(410,200)[tl]{\small 10 fb$^{-1}$}
}\end{comment}\end{picture}
}
\caption[f11]{\small Contours in the \susy\ parameter space
indicating the required luminosities
for distinguishing a \sel\ signal
from the \sm\ background
at a $3\sigma$ confidence level.
The cuts (\ref{egcuts}) are included.
Results are shown for 10, 20, 50 and 100 fb$^{-1}$
of integrated luminosities.
The shaded area
is explorable in $e^{\pm}e^-$ collisions.
The dotted line indicates the kinematic boundary in \eg\ scattering.
The photon beam is assumed to have the energy spectrum (\ref{e3}).}
\label{f11}
\end{figure}

The situation can be further improved
by using left-polarized electron beams
to cut down the $W^-$ background (\ref{ez}).
If the photon beam is also polarized
(by polarizing the electron and laser beams)
one can similarly reduce
the $Z^0$ background \cite{LC19}.
A thorough study of these suggestions
has still to be performed.

\section{$\gamma\gamma$ Collisions}

Sleptons can be pair-produced
by colliding photon beams,
as depicted in Fig.~\ref{f12}.
The energy dependence of the \xs\ is also illustrated
in Fig.~\ref{f12}.
The threshold behaviour is washed out here
by the broad energy distribution (\ref{e3})
of the colliding photons.
Therefore,
this reaction cannot yield much information
on the mass of the \sel.
Moreover,
since the maximum $\gamma\gamma$ energy
is almost 20 \%\ lower than the $e^+e^-$ or $e^-e^-$ energies,
this option is not competitive with the other
\lc\ modes for discovering \co s or sleptons.

On the other hand,
$\gamma\gamma$ collisions posses
the great advantage
that there is no model dependence at the production level.
This is because photons couple solely to electric charges,
which are fixed even for hypothetical particles.
One can, therefore, investigate
the decay properties of the \sel\ separately in a clean way.
This may yield important information on the gaugino-higgsino sector.
The signal we focus on here

\medskip
\centerline{\fbox{$\gamma\gamma \to \tilde e^-\tilde e^- \to e^+e^-+\mpT$}}

has been analysed in \cite{wo18}.
The same analysis can be applied
to the $\mu$ signal
and, with some restrictions, also to the $\tau$ signal.
Besides sleptons also
charginos can be pair-produced in \pp\ collisions and
decay into the final states considered above.
However,
the \xs\ for an observable signal is much lower.

\begin{figure}[htb]
\begin{center}
\begin{picture}(250,500)(0,-100)
{\tt
	The figures can be obtained via anonymous ftp from 129.187.198.1
	in preprints/lc93/susy.ps .
}
\begin{comment}{
\Photon(00,100)(37.5,105){2}{5}
\Photon(00,150)(37.5,145){2}{5}
\Vertex(37.5,105){.5}
\DashLine(37.5,105)(37.5,145){5}
\Vertex(37.5,145){.5}
\DashLine(37.5,105)(75,100){5}
\DashLine(37.5,145)(75,150){5}
\Text( -5,200)[r]{$\gamma$}
\Text( -5,300)[r]{$\gamma$}
\Text( 80,250)[l]{$\tilde{e}$}
\Text(155,200)[l]{$\tilde{e}^-$}
\Text(155,300)[l]{$\tilde{e}^+$}
\Text(75,150)[c]{+}
\Photon(0,00)(37.5,25){2}{5}
\Photon(37.5,25)(0,50){2}{5}
\Vertex(37.5,25){.5}
\DashLine(37.5,25)(75,0){5}
\DashLine(37.5,25)(74,50){5}
\Text(-5,0)[r]{$\gamma$}
\Text(-5,100)[r]{$\gamma$}
\Text(155,100)[l]{$\tilde e^+$}
\Text(155,0)[l]{$\tilde e^-$}
}\end{comment}\end{picture}
\begin{picture}(554,504)(0,0)
\begin{comment}{
\put(0,0){\strut\bBoxedEPSF{/user/frank/Papeles/LC93/gg.ps}}
\Text(65,460)[tr]{\small$\sigma$[pb]}
\Text(520,19)[tr]{\small$\sqrt{s_{ee}}$[GeV]}
\Text(300,300)[tl]{\fbox{\small $\gamma\gamma \to \tilde e^+\tilde e^-$}}
%\Text(300,225)[tl]{\small $\tan\beta = 10$}
%\Text(300,200)[tl]{\small $\mu = -300$ GeV}
%\Text(300,175)[tl]{\small $M_2 = 200$ GeV}
\Text(300,225)[tl]{\small $m_{\tilde\ell} = 150$ GeV}
}\end{comment}\end{picture}
\end{center}
\caption[f12]{\small Lowest order Feynman diagrams
describing \sel\ production in $\gamma\gamma$ collisions
and typical energy dependence of the corresponding \xs.
For the photon beams, the energy spectrum (\ref{e3})
is assumed.}
\label{f12}
\end{figure}

The only potentially dangerous \sm\ background
to the $\pp\to e^+e^-+\mpT$ signal
is $W$ pair production
{\arraycolsep0cm
\renewcommand{\arraystretch}{0}
\begin{equation}
        \gamma\gamma \to
        \begin{array}[t]{ll}
                W^+&W^- \qquad . \\
                      &\hra e^-\bar\nu_e\\
                \makebox[0cm][l]{$\hra e^+\nu_e$}
        \end{array}
\label{enwpm}
\end{equation}
}
However,
the signal to background ratio can be substantially enhanced
by imposing the \trm, rapidity and acoplanarity cuts:
\begin{equation}
\pT(e^+)\pT(e^-) > m_W^2
\quad , \quad
|\eta(e^\pm)| < 1
\quad , \quad
||\phi(e^+)-\phi(e^-)|-180^\circ| > 2^\circ
\label{e102}\ .
\end{equation}
The last cut is imposed to eliminate the
largest but simplest background from $e^+e^-$-pair production
$\gamma\gamma \to e^+e^-(\gamma)$.
The background can be even further reduced
without affecting the signal
if the masses of the \sel\ and \No\ are already known
(for example,
from the pair-production threshold in \ee\ collisions
and Eq.~(\ref{e100})).
In that case,
one can reject all events in which the electron energies
lie outside the boundaries
\begin{equation}
{\sqrt{s_{\gamma\gamma}^{\rm max}}\over4}
\left[1-{m^2_{\No}\over m^2_{\Sel}}\right]
\left[1\pm\sqrt{1-{4m^2_{\Sel}\over s_{\gamma\gamma}^{\rm max}}}\right]\ ,
\label{eny}
\end{equation}
where $\sqrt{s_{\gamma\gamma}^{\rm max}}
=2(\sqrt{2}-1)\sqrt{s_{ee}} \approx .83\sqrt{s_{ee}}$
is the maximum attainable \cm\ energy
in the photon-photon collision.

\begin{figure}[htb]
\centerline{
\begin{picture}(250,500)(0,-350)
{\tt
	The figures can be obtained via anonymous ftp from 129.187.198.1
	in preprints/lc93/susy.ps .
}
\begin{comment}{
\Text(0,0)[tl]{\small \fbox{$\gamma\gamma\to e^+e^-+\mpT$}}
\Text(0,-75)[tl]{\small $\sqrt{s_{ee}} = 1000$ GeV}
\Text(0,-100)[tl]{\small $\tan\beta = 4$}
\Text(0,-125)[tl]{\small $m_{\tilde\ell} = 300$ GeV}
}\end{comment}\end{picture}
\begin{picture}(554,504)(0,0)
\begin{comment}{
\put(0,0){\strut\bBoxedEPSF{gg.hacked.ps}}
\put( 60.8,468.9){\makebox(0,0)[tr]{\small$M_2$[GeV]}}
\put(480.8, 19.1){\makebox(0,0)[tr]{\small$\mu$[GeV]}}
\put(120,190){\makebox(0,0)[bl]{\small$ 1$ fb$^{-1}$}}
\put(120,300){\makebox(0,0)[bl]{\small$ 10$ fb$^{-1}$}}
\put(462,196){\makebox(0,0)[br]{\small$ 1$ fb$^{-1}$}}
\put(462,305){\makebox(0,0)[br]{\small$ 10$ fb$^{-1}$}}
\Text(112.2,426.9)[lb]{\scriptsize unphysical}
\Text(389.6,426.9)[lb]{\scriptsize unphysical}
}\end{comment}\end{picture}
}
\caption[f14]{\small Contours in the \susy\ parameter space
indicating the luminosities required
for distinguishing a \sel\ signal
from the \sm\ background
at a $3\sigma$ confidence level.
The cuts (\ref{e102},\ref{eny}) are included.
For the photon beams, the energy spectrum (\ref{e3}) is assumed.
Results are shown for 1, 2, 5 and 10 fb$^{-1}$
of integrated luminosities.}
\label{f14}
\end{figure}

The explorable parameter space is outlined in Fig.~\ref{f14},
where contours
along which the signal exceeds the backgrounds statistical fluctuations
by more than 3 standard deviations
($n_{SUSY}>3\sqrt{n_{SM}}$)
are depicted.

\section{Summary and Conclusions}

We have studied the production and decay of \sel s and \co s
in \ee, \eg\ and \pp\ collisions.
Our results can be summarized as follows:

\begin{description}

\item[\underline{\boldmath$e^-e^-$ Collisions:}]
\item{\boldmath$\oplus$} Privileged mode for {\bf discovery},
        because of inherently low \sm\ backgrounds.
        The signals are more striking than in $e^+e^-$ collisions.
\item{\boldmath$\oplus$} Cascade decays of the \sel\
        can be observed and studied,
        because no high \trm\ cuts are needed
        to enhance the signal to background ratio.
\item{\boldmath$\ominus$} Difficult to disentangle the effects of
        masses, mixings and branching ratios.

\item[\underline{\boldmath$e^-\gamma$ Collisions:}]
\item{\boldmath$\oplus$} If \sel s are too heavy to be pair-produced,
        this is the only potential {\bf discovery} mode.
\item{\boldmath$\ominus$} High luminosities are required.
        Polarization of the electron and photon beams helps.

\item[\underline{\boldmath$\gamma\gamma$ Collisions:}]
\item{\boldmath$\oplus$} Ideal mode for {\bf study}
        of gaugino/higgsino parameters which
        determine the decay properties of sleptons.
\item{\boldmath$\ominus$} Reach in mass restricted due to the
        reduced energies of the photon beams relative to the
        initial energy of the electron beams.
\end{description}

Comparing \ep\ and \ee\ collisions
it should be noted that while the production cross sections
are similar in the two reactions, the latter
suffers much less from the \sm\ backgrounds.
Hence, \ee\ collisions are expected to provide the cleanest
sample of \susic\ events.

To conclude, we emphasize that the possibility of experimenting
with (polarized)
$e^-e^-$, $e^-\gamma$ and $\gamma\gamma$ beams
is a unique feature of linear colliders,
whose usefulness
is by no means restricted to the search and study of \susic\ particles.
On the contrary,
these options have been shown to also provide powerful tests of
the \sm\ \cite{LC7},
anomalous gauge couplings \cite{LC28},
gauge extensions of the \sm\ \cite{F1},
Majorana neutrinos \cite{H1},
and many other interesting speculations.
With such unique possibilities in mind
one would clearly favor
accelerator and detector designs
which are not fundamentally preventing these options.

\bigskip
\bigskip

Part of the \xs s we used have been checked with the help
of the computer program CompHEP \cite{CHEP1}.
We are very much indebted to Edward Boos and Michael Dubinin
for having provided us with this software.
This work was partially supported by the German Federal
Ministry for Research and Technology
under contract No. 05 6MU93P
and by the CED Science Project No. SCI-CT 91-0729.

\end{document}